\documentclass[prb,twocolumn,showpacs,floatfix]{revtex4}

  \usepackage{graphicx}

\newcommand{\bra}[1]{\langle{#1}|}
\newcommand{\ket}[1]{|{#1}\rangle}

\newcommand{\op}[1]{\hat{#1}}

\newcommand{\set}[1]{\left\{{#1}\right\}}

\newlength{\figwidth}
\begin{document}

\title{Experimental realization of the one qubit Deutsch-Jozsa
algorithm in a quantum dot}

\author{P. Bianucci}
\author{A. Muller}
\author{C. K. Shih}
\email{shih@physics.utexas.edu}
\affiliation{Physics Department, The University of Texas at Austin, Austin, 
             Texas 78712}
\author{Q. Q. Wang}
\altaffiliation{Department of Physics, Wuhan University, Wuhan 430072, 
                P. R. China}
\author{Q. K. Xue}
\affiliation{International Center for Quantum Structures, Institute of Physics,
             The Chinese Academy of Sciences, Beijing 100080, P. R. China}
\author{C. Piermarocchi}
\affiliation{Department of Physics and Astronomy, Michigan State University, 
             East Lansing, Michigan 48824-2320}

\date{\today}

\begin{abstract}
We perform quantum interference experiments on a single self-assembled
semiconductor quantum dot.  The presence or absence of a single exciton in the
dot provides a qubit that we control with femtosecond time resolution. 
We combine a set of quantum operations to realize the single-qubit
Deutsch-Jozsa algorithm. The results show the feasibility of single qubit
quantum logic in a semiconductor quantum dot using ultrafast optical control. 
\end{abstract}

\pacs{78.55.Cr,71.35.-y,03.67.Lx}

\maketitle

Time-resolved optical spectroscopy in semiconductor quantum dots has recently
progressed toward the full quantum control of excitons trapped inside a single
dot.\cite{BonadeoSCI98,StievaterPRL01,ZrennerNAT02,HtoonPRL02} These advances
have stimulated proposals to use excitons in quantum dots as quantum bits
\cite{TroianiPRB00,ChenPRL01-2,BiolattiPRB02} for implementation of quantum
computing.  Very recently, the ability to operate a two-qubit gate using exciton
and biexciton states was demonstrated in a single quantum dot.\cite{LiSCI03}
These achievements represent a step toward an all-optical implementation of
quantum computing using excitonic qubits.  The first algorithm that comes to
mind in order to check the feasibility of quantum computation in this context is
the Deutsch-Jozsa (DJ) algorithm.\cite{DeutschPRS92} This algorithm is one of
the simplest quantum algorithms that provides an exponential speed-up with
respect to classical algorithms.  As such, it has been extensively studied and
has been used in experimental demonstrations of simple quantum computation in a
variety of systems.  \cite{ChuangNAT98,LindenCPL98,GuldeNAT03} In this Rapid
Communication we report the experimental realization of the DJ algorithm for a
single qubit using an optimized version of the algorithm\cite{CollinsPRA98}.

The Deutsch problem \cite{DeutschPRS92} involves global properties of binary
functions on a subset of the natural numbers.  Given a natural number $N$, we
can define a set called $X_N$ with all the natural numbers that can be
represented with $N$ bits.  A binary function $f:X_N \to \set{0,1}$ is called
balanced if it returns $0$ for exactly half of the elements of $X_N$ and $1$ for
the other half.  Given a function that is either balanced or constant, the
Deutsch problem consists of finding out which type it is.  A general classical
algorithm requires evaluating the function on more than half of the elements,
requiring at least $2^{N-1}+1$ evaluations.  This causes the classical run time
to grow exponentially with the input size.  The Deutsch-Jozsa algorithm provides
a way to solve the Deutsch problem on a quantum computer using a quantum
subroutine that evaluates $f$.  The problem and its solution provide an example
of Oracle-based quantum computation.\cite{BerthiaumeJMO94,BennettSJC97} It is
assumed that a quantum subroutine or Oracle contains the information about the
unknown function.  The algorithm gives a recipe on how to prepare (encoding) and
read out (decoding) the qubit in an efficient way.  In an experimental
demonstration, we have not only to implement the algorithm (encoding and
decoding operations), but we also have to build the Oracle.  The specific
structure of the Oracle, encoding and decoding is not unique and several
versions can be found in the
literature.\cite{DeutschPRS92,ChuangPRA95,ClevePRS98,CollinsPRA98}.  The one we
are using here\cite{CollinsPRA98} allows us to implement the $N$=1 case with a
single qubit.
\begin{figure}
  \centering
  \includegraphics[width=\figwidth,keepaspectratio,clip]{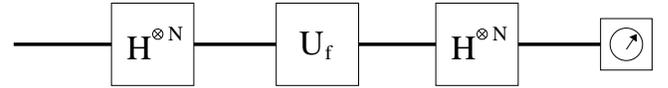}
  \caption{Optimized version of the Deutsch-Jozsa algorithm.}
  \label{fig:qcirc}
\end{figure}
Figure \ref{fig:qcirc} shows a quantum circuit depiction of the algorithm. 
This circuit uses the following quantum transformations:
\begin{enumerate}
  \item A Hadamard transformation independently applied to each qubit, 
        $\op{H}^{\otimes N} = \op{H} \otimes \ldots \otimes \op{H}$. A
        single qubit transformation is represented by
    \begin{equation}
        \op{H} = \frac{1}{\sqrt{2}}\left[
                                     \begin{array}{rr}
                                        1 &  1 \\
                                        1 & -1 \\
                                     \end{array}
                 \right].
    \end{equation} 
  \item A $f$-controlled gate, whose operation is defined as
     \begin{equation}
     \op{U}_f \ket{x} = (-1)^{f(x)} \ket{x}.
     \end{equation}
\end{enumerate}
The final step in the algorithm measures the expectation value of the
$\ket{0}\bra{0}$ operator.  This expectation value for a constant function will
be equal to $1$ while for a balanced function it will be equal to $0$.
When $N$=1 there are only four possible functions $f_j:\set{0,1} \to
\set{0,1}$:
\begin{eqnarray}
  f_1(x) = 0, \\ 
  f_2(x) = 1, \\
  f_3(x) = x, \\
 \mbox{and}~ f_4(x) = 1-x.
\end{eqnarray}
Of these four, $f_1$ and $f_2$ are constant while $f_3$ and $f_4$ are balanced. 
The explicit matrix forms of the $\op{U}_f$ operators are:
\begin{equation}
  \op{U}_{f_1} = \left[
                   \begin{array}{rr}
                     1 & 0 \\
                     0 & 1 \\
                   \end{array} 
                 \right]
               = \op{I}, 
  \quad 
  \op{U}_{f_2} = -\left[ 
                    \begin{array}{rr}
                     1 & 0 \\
                     0 & 1 \\
                   \end{array}
                 \right]
               = -\op{I},
\end{equation}
\begin{equation}
  \op{U}_{f_3} = \left[
                   \begin{array}{rr}
                     1 &  0 \\
                     0 & -1 \\
                   \end{array}
                 \right]
               = \op{\sigma_z}, 
  ~\mbox{and}~
  \op{U}_{f_4} = -\left[
                   \begin{array}{rr}
                     1 &  0 \\
                     0 & -1 \\
                   \end{array}
                 \right]
               = -\op{\sigma_z}.
\end{equation}
We can see that the balanced functions share the same $f$-controlled operator
except for a global phase.  This is also true for the constant functions.
If the qubit is initially in the state $\ket{0}$, the encoding transformation
consists in one Hadamard operation that transforms the qubit to
\begin{equation}
  \frac{1}{\sqrt{2}}(\ket{0}+\ket{1}).
  \label{out1}
\end{equation}
By applying  $\op{U}_{f_j}$ to the state in Eq.\ \ref{out1} we obtain
\begin{equation}
  \op{U}_{f_j}\frac{1}{\sqrt{2}}(\ket{0}+\ket{1}) =
  \frac{1}{\sqrt{2}}[(-1)^{f_{j}(0)}\ket{0}+(-1)^{f_{j}(1)}\ket{1}].
\end{equation}
For a constant function this gives
\begin{equation}
  (-1)^{f_{j}(0)}\frac{1}{\sqrt{2}}(\ket{0}+\ket{1}),
\end{equation}
while for a balanced function we get
\begin{equation}
  (-1)^{f_{j}(0)}\frac{1}{\sqrt{2}}(\ket{0}-\ket{1}).
\end{equation}
As a decoding procedure, we apply again the Hadamard transformation. We obtain 
\begin{equation}
  (-1)^{f_{j}(0)}\ket{0}
\end{equation} 
for a constant function, and 
\begin{equation}
  (-1)^{f_{j}(0)}\ket{1}
\end{equation}
for a balanced function.  Therefore, by measuring one of the two states, one can
decide in a deterministic way to which class $f$ belongs.  We remark that if we
were to obtain an answer using only classical operations, we would need to
evaluate the unknown $f$ function twice, obtaining both $f(0)$ and $f(1)$ and
comparing them.  Conversely, the described quantum procedure only requires one
call of the quantum subroutine $\op{U}_f$ is needed.  Therefore the $N$=1 case
of the DJ already shows that the quantum algorithm outperforms its classical
counterpart by a factor of two in the number of evaluations. 

\begin{figure}
  \centering
  \includegraphics[clip]{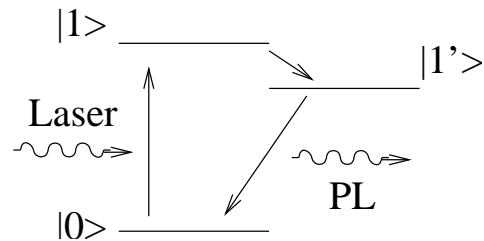}
  \caption{Quantum level structure. The excitonic ground state and first 
           excited state are labeled $\ket{1'}$ and $\ket{1}$ respectively. The
           state $\ket{0}$ corresponds to the absence of an exciton in the 
           quantum dot.}
  \label{fig:levels}
\end{figure}

We have been able to implement the single-qubit Deutsch-Jozsa algorithm
discussed above using the excitonic states of a self-assembled InGaAs quantum
dot as a qubit.  The level scheme we used is depicted in Fig.\ \ref{fig:levels}. 
The absence of an exciton is taken as the $\ket{0}$ state of the qubit, while
the first excited excitonic state is taken as $\ket{1}$.  The $\ket{1}$ state
population is monitored via a non-radiative transition to the exciton ground
state (labeled as $\ket{1'}$) whose radiative recombination is recorded using a
micro-photoluminescence
setup.\cite{HtoonPRB99,HtoonAPL00,HtoonPRB01,HtoonPRL02,MullerAPL04}

We will use two different unitary transformations to realize the Deutsch-Jozsa
algorithm:  a $\frac{\pi}{2}$ single qubit rotation and a phase shift.  The
corresponding explicit matrix forms are:
\begin{equation}
  \op{U}_{\frac{\pi}{2}} = \frac{1}{\sqrt{2}}\left[
                   \begin{array}{rr}
                     1 & -1 \\
                     1 &  1 \\
                   \end{array} 
                 \right]
\end{equation}
and
\begin{equation}
  \op{U}(\phi) = \left[ 
                    \begin{array}{rr}
                     e^{-i\frac{\phi}{2}} &         0           \\
                     0 & e^{i\frac{\phi}{2}}\\
                   \end{array}
                 \right].
\end{equation}
The single qubit rotation is realized by a $\pi/2$ pulse resonant with the
$\ket{0}$ to $\ket{1}$ transition.  We use the rotating wave approximation
and the qubit is defined in the rotating frame.  The phase gate $\op{U}(\phi)$
is realized by controlling the phase of the optical pulses with respect to the
first pulse which is used as a reference.  This is achieved experimentally by a
piezoelectric translation stage that controls the phase locking between the
pulses.  By choosing specific values for $\phi$, $\op{U}(\phi)$ becomes
equivalent to the $f$-controlled operators, as shown in Table I.  In this
version of the algorithm, the Oracle distinguishes the operations within the
same class only by a global phase in the single qubit space.  We can always
think about an additional reference qubit in the Oracle to make this phase
physically measurable.  However, this reference qubit will never come into play
in the real algorithm since it is part of the internal structure of the Oracle.


Notice that although $\op{U}_{\frac{\pi}{2}}$ and $\op{H}$ behave in a similar
way, they are not the same operator.  It is easy to show that the only effect
of this change is that the interpretation of the final result has to exchange
balanced with constant functions.  We can think about the quantum evolution of
the qubit during the algorithm using the picture of a pseudo-spin in the Bloch
sphere.  The first pulse corresponds to an effective magnetic field in the
$+{\bf y}$ direction that brings the pseudo-spin from $-{\bf z}$ to the $-{\bf
x}$ direction.  The phase shift corresponds to a rotation of the pseudo-spin
around the ${\bf z}$ axis of multiples of $\pi$.  The second pulse will bring
the pseudo-spin back to $-{\bf z}$ in the case of a balanced function (by
destructive interference), and to $+{\bf z}$ in the case of a constant
function.  In this picture the $N$=1 Deutsch algorithm shows clearly its
equivalence to a Mach-Zehnder interferometer experiment.\cite{ClevePRS98} 

The sample consisted of In$_{0.5}$Ga$_{0.5}$As MBE grown self-assembled quantum
dots, kept at a temperature of 5 K inside a continuous flow liquid helium
cryostat.  The quantum dots were resonantly excited with pulses from a
mode-locked Ti:Sa laser.  The pulses were linearly polarized in a way to make
sure only one state out of an anisotropy induced doublet was
excited.\cite{MullerAPL04} By using a spectrometer combined with a
two-dimensional liquid nitrogen cooled charge-coupled device (CCD) array
detector, we were able to detect the integrated photoluminescence signals of
many quantum dots at the same time.\cite{HtoonAPL00} This enabled us to search
for a quantum dot with a large enough dipole moment (and thus a good
signal-to-noise ratio) and a dephasing time larger than 20 ps for the excited
state, which is the case for about 1\% of the dots.  We did not select any
specific polarization at the detection.

The use of the excitonic ground state photoluminescence as the means of
detection prevented us from being able to use this state as the $\ket{1}$ state
of our qubit. This entailed a severe decrease in the dephasing time of the
qubit, as the non-radiative decay from the excited state to the exciton ground
state (necessary for our detection scheme to work) puts an upper bound in the
coherence time of the exciton\footnote{Other mechanisms might further reduce the
coherence time.}.  This upper bound is significant, since measured dephasing
times for excitonic ground states are in the order of hundreds of
picoseconds\cite{BorriPRL01,BirkedalPRL01,BayerPRB02} while those for carefully 
chosen excited states (i.e. no further than approximately 20 meV apart from the
corresponding ground state) range in the tens of picoseconds.\cite{HtoonPRB01}

The actual implementation of the algorithm was similar to that of standard wave
packet interferometry measurements,\cite{BonadeoSCI98,KamadaPRL01} but in the
nonlinear excitation regime.\cite{HtoonPRL02} In order to establish the
appropriate excitation intensity for a $\frac{\pi}{2}$ pulse, we first recorded
Rabi Oscillations of the excited state.\cite{KamadaPRL01,HtoonPRL02} We also
performed a low intensity wave packet interferometry measurement to estimate the
dephasing time of the quantum dot.\cite{BonadeoSCI98,KamadaPRL01} In that
experiment, the photoluminescence signal is proportional to the wavefunction
autocorrelation.  By fitting the decay of the autocorrelation signal with an
exponential function we were able to measure the dephasing time of the exciton
in the dot, obtaining 40 ps as a result. 

In the main experiment, the time delay between two identical resonant
$\frac{\pi}{2}$ laser pulses (approximately 5 ps long) was scanned while
simultaneously recording the photoluminescence.  A mechanical translation stage
controlled the coarse delay between the two pulses while a piezoelectric stage
changed the fine delay.  The fine delay is used to control the phase shift of
the second pulse with respect to the first one.  It can be mapped to the
relative phase by the relation $\phi=\omega_0 \tau$, where $\hbar\omega_0$ is
the laser energy, and has been calibrated by performing wavepacket
interferometry at low intensity on the quantum dot, keeping the mechanical stage
at a fixed position. 

\begin{figure}
  \centering
  \includegraphics[width=\figwidth,keepaspectratio,clip]{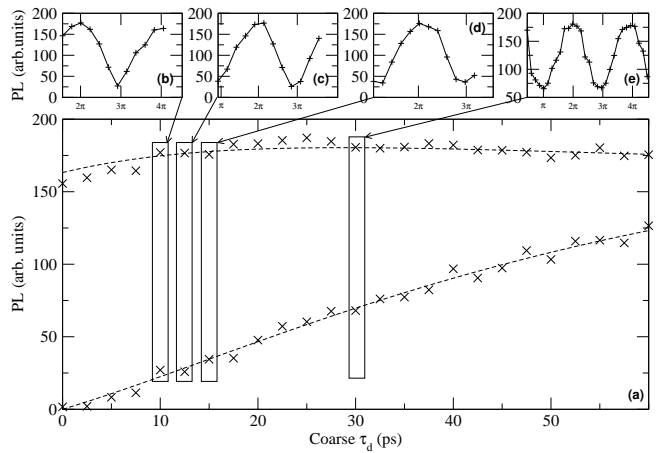}
  \caption{Central plot: Envelope of the photoluminescence (PL) as a function
           of the coarse pulse delay. PL signals as a function of the phase
           difference between the two pulses are shown in the insets.}
  \label{fig:data}
\end{figure}

The encoding and decoding consist of the preparation of the two pulses with the
same phase.  We can imagine that the Oracle controls the fine delay knob, and,
by changing the relative phase, determines which one of the four functions is
being implemented.  Figure \ref{fig:data}a shows the intensity of the detected
photoluminescence as a function of the coarse delay between the two pulses.  The
lower and upper signals correspond to constructive and destructive interference
depending on the relative phase of the two pulses.  The contrast between the
maxima and minima of the signal decreases as the delay between the pulses
approaches the dephasing time of the dot (40 ps), leading to lower fidelities. 
Figures \ref{fig:data}b-e describe the detailed behavior of the signal for
various values of the phase difference between the two pulses.

We can now interpret this result in terms of the DJ quantum algorithm.  As
expected, the maximum population at $\ket{1}$ (that is maximum
photoluminescence) occurs for even numbers of $\pi$ in the relative phase
between the two pulses, corresponding to the constant quantum subroutines
$\op{U}_{f_{1,2}}$.  On the other hand, minima occur for odd numbers of $\pi$ in
the phase shift between the two pulses, corresponding to the balanced quantum
subroutines $\op{U}_{f_{3,4}}$.  The probability of successfully solving the
problem is related to the contrast of the maxima and minima in the interference
process.  We remark that the first three insets in Fig.\ \ref{fig:data} (all
with a delay between the pulses between 10 and 20 ps) have a contrast of the
order of 75\%.  This implies a fidelity for the quantum operations comparable to
other similar implementations.\cite{LiSCI03} The fidelity is mainly limited by
the dephasing time of the excited excitonic state of the quantum dot.  Making
the coarse delay between the pulses as short as possible gives the best fidelity
(as can be seen in Fig.\ \ref{fig:data}), but this delay must be no shorter than
twice the excitation pulse width, so that any optical interference arising out of
the overlap of the two pulses is negligible.  Also, a detection scheme able to
resonantly excite and then measure the exciton ground state would allow for much
larger fidelities, due to the increased coherence times.

\begin{table}[ht]
 \begin{ruledtabular}
  \begin{tabular}{cc}                                     \hline
    Experimental phase shift     &  Operation          \\ \hline
    $4n\pi$	&  $\op{U}_{f_{1}}$	\\ \hline
    $\pi+4n\pi$	&  $-i\op{U}_{f_{3}}$	\\ \hline
    $2\pi+4n\pi$&  $\op{U}_{f_{2}}$	\\ \hline
    $3\pi+4n\pi$&  $-i\op{U}_{f_{4}}$	\\ \hline
  \end{tabular}
  \label{tbl:f}
  \caption{Experimental phase shift and their implemented operations}
 \end{ruledtabular}
\end{table}

By using an interferometric set-up on an excitonic qubit system, we have been
able to implement the single-qubit Deutsch-Jozsa algorithm.  Although the
1-qubit version of the algorithm does not show all the features of Quantum
Computing (in particular entanglement), it is an experimental demonstration of
simple quantum computation, including superpositions and interference, in
a solid state system. 

\begin{acknowledgments}
We would like to thank John Robertson for proof-reading the article and for 
his style corrections.
This work was supported by NSF-NIRT (DMR-0210383), NSF-FRG (DMR-0306239), 
NSF-ITR (DMR-0312491), the Texas Advanced Technology program, and the W.M. 
Keck Foundation.
\end{acknowledgments}

\bibliography{1qdj}

\end{document}